\documentclass[manuscript]{aastex}

\def \msun   {\hbox{M$_\odot$}}

\shorttitle{Detectability of Exoplanets in BPMG using GPI}
\shortauthors{Kataria \& Simon}

\begin{document}

\title{Detectability of Exoplanets in the Beta Pic Moving \\ Group with the Gemini Planet Imager}

\author{Tiffany Kataria\altaffilmark{1} and Michal Simon}
\affil{Department of Physics and Astronomy, Stony Brook University,
    Stony Brook, NY 11794}

\altaffiltext{1}{Present address: Lunar and Planetary Laboratory, University of Arizona}

\begin{abstract}

We model the detectability of exoplanets around stars in the Beta Pic Moving Group 
(BPMG) using the Gemini Planet Imager (GPI), a coronagraphic instrument designed to 
detect companions by imaging.  Members of the BPMG are considered promising targets for 
exoplanet searches because of their youth ($\sim 12$ MY) and proximity (median distance $\sim 35$ pc).  
We wrote a modeling 
procedure to generate hypothetical companions of given mass, age, eccentricity, 
and semi-major axis, and place them around BPMG members that fall within the V-band range
of the GPI.  We count as possible detections companions lying within the GPI's field of view 
and H-band fluxes that have a host-companion flux ratio placing them within its sensitivity.  
The fraction of companions that 
could be detected depends on their brightness at 12 Myr, and hence formation  mechanism, and 
on their distribution of semi-major axes.  We used brightness models for formation by disk instability
and core-accretion.  We considered the two extreme cases of the semi-major axis distribution - the 
log-normal distribution of the nearby F and G type stars and a power-law distribution
indicated by the exoplanets detected by the radial velocity technique.  We find that the GPI 
could detect exoplanets of all the F and G spectral type stars in the BPMG sample with a probability 
that depends on the brightness model and semi-major axis distribution.  At spectral type K to M1, 
exoplanet detectability depends on brightness and hence distance of the host star.  
GPI will be able to detect the companions of M stars later than M1 only if they are closer 
than 10 pc. Of the four A stars in BPMG sample, only one has V-band brightness in the range 
of GPI; the others are too bright.   

\end{abstract}


\section{Introduction}

The primary goal of the Gemini Planet Imager (GPI), a coronagraphic instrument under construction for the Gemini
Observatory (Graham et al. 2007; see also {\it Future Instrumentation} at www.gemini.edu), is the detection of    
exoplanets directly by imaging.  Young stars will be important targets for GPI science because their
exoplanets  are expected to dim as they age and cool.
Over the past 20 years astronomers have identified groups of relatively nearby
young stars distinguished by their space motion through the galaxy,
hence the name “Nearby Young Moving Groups” (NYMGs).  The TWA group at median distance
56 pc is the youngest, $\sim 8$ MY, and the $\beta$ Pic and AB Dor groups are the nearest 
with median distances 35 and 30 pc, respectively (Zuckerman and Song, 2004; Torres et al. 2006; 
Fern\'andez et al. 2008).  
Their ages are about 12 MY for $\beta$ Pic and 30-100 MY for AB Dor (Fern\'andez et al. 2008). 
By analyzing the kinematics of the NYMGs, Fern\'andez et al. (2008) showed 
that they share a common origin at the edges of the giant molecular cloud that formed the Sco-Cen 
star forming region. On this scenario, the supernova that heated the Local Bubble also triggered 
the formation of the stars in the NYMGs.  

Owing to their youth and proximity stars in the NYMGs are promising targets for exoplanet searches 
by high resolution and high contrast imaging. With detectable contrast ratios of $10^7$, 
the GPI is designed to image brown dwarfs and exoplanets that may surround a star.  
Astronomers can then spectrally analyze the composition of
their atmospheres, which has further implications in astrobiology and stellar and planetary evolution
(eg. Oppenheimer and Hinkley, 2009).  L\'epine and Simon (2009) described a pilot program to identify
new low-mass probable members of the $\beta$ Pic moving group (BPMG). Schlieder et al. (2009) 
are continuing the program and expanding it to include the AB Dor moving group.  Our goals
in the work described here were to evaluate the detectability  of exoplanets in the BPMG
and hence to provide guidance for observing programs of the BPMG, and other NYMGs, as new 
members are discovered.

\section{Detectability of Brown Dwarfs and Exoplanets in the BPMG}

\subsection{GPI Field of View and Sensitivity}

Detectability of companions using the GPI is determined by its field of view (FOV) and
by its sensitivity to the flux of faint companions near the much brighter host star. 
The $0.22''$ radius of the occulting disc of the coronagraph sets the inner boundary 
of the FOV.  Observations are made with the host star image centered behind the 
occulting disk.  The outer boundary is a square of dimensions $ 2.8'' \times 2.8''$
centered on the occulting disk (Graham et al. 2007).  At the 35 pc median distance of the 
BPMG members,  the FOV of the GPI encompasses host-companion separations $\sim 7$ to 
50 AU.   For a 1 \msun ~ host, this implies orbital periods  $\sim 18$ to 350 y. 
 
The most important parameter determining the detectability of a companion within
the FOV is the ratio of its brightness to that of the host star.   The detectable brightness
ratio, or contrast, in turn depends on the quality of the wavefront correction achieved
by the telescope's adaptive optics (AO) system.  Thus, the detectable contrast depends
on the brightness of the host star at V, the band at which the AO system operates, and the 
wavelength at which the companion is observed, usually the H-band for the GPI.
The contrast curves (Figure 1) are expressed as the 5-$\sigma$ H-band magnitude difference 
between the companion and host star as a function of their angular separation that is 
detectable in 1 hour of observation (J. Graham, private communication).  The detectable 
contrast assumes installation at Gemini-South and that speckle noise dominates the images.  
The curves are given at 5-$\sigma$ values of the host star brightness, V= 5 to 9 mag.

\subsection{Modeling the BPMG Sample}

Table 1 lists the 29 published members of the BPMG  (Fern\'andez et al 2008; L\'epine and Simon 
2009).  The J, H, and K-band magnitudes are from the {\it 2MASS} survey.  We evaluated the 
detectability of possible companions  for BPMG members in the brightness range at which 
the contrast curves are given and extropolated them to V=10 in order to include HIP 29964. 
Our analysis thus included 15 BPMG stars\footnote{The A spectral type stars HIP 79881 
and HD 172555 with V=4.9 and 4.8 are slightly brighter than the V=5 sensitivity curve in 
Figure 1.  The nature of the companions for these  stars should be very similar to that 
for the 5.0 magnitude A0V star $\eta$Tel.}; these are identified with an asterisk in Table 1, column 1.
We needed to estimate the host star mass in order to model the companion's motion.  
To accomplish this, we used the spectral type listed by Fern\'andez et al. 
2008), Kenyon and Hartmann's (1995) spectral type to effective temperature scale, 
and estimated the mass with Siess' web-based isochrone calculation 
tool\footnote{http://www-astro.ulb.ac.be/~siess/server/iso.html} at age 10 MY (Siess et al. 2000).  

According to current views, planets are thought to form in circumstellar disks by one of 
two mechanisms: by core-accretion (Lissauer  and Stevenson 2007) or by disk instability 
(Durisen et al. 2007).  These are dubbed the ``cold start'' and ``hot start'' mechanisms 
(eg. Fortney et al. 2008, F08; Nielsen and Close 2009).  F08 calculated  the IR 
luminosities of 1-10 $M_{J}$ companions at 10 MY age for the core-accretion and 
``hot start'' formation models (our Table 2).  Baraffe et al. (1998, 2003) also 
calculated the IR luminosities of  companions for the hot start scenario (our Table 3); 
their H-band luminosities for 1-10 $M_{J}$ exoplanets are similar to F08's values but 
slightly brighter. In our modeling we use F08's results for the core-accretion luminosities 
and   Baraffe et al's results for the disk instability luminosities.  The latter 
cover a larger mass range than F08's disk instability luminosities.

The exoplanet semi-major axis distribution at separations falling within the GPI's FOV, 
7 to 50 AU for host star at 35 pc, is unknown at present.  It will not be known until
many more exoplanetary systems like those associated with Fomalhaut and HR 8799 (Kalas et 
al 2008; Marois et al 2008) are identified.  We use, therefore, two distributions for our 
analysis that may represent the possible extremes.  The first is that of stellar 
companions orbiting nearby main sequence F and G stars (Duquennoy \& Mayor 1991; 
hereafter DM91). We binned the distribution at $\Delta log~P(d)=  1$ (Figure 2) and 
drew values of the period randomly according to frequencies indicated.   The short 
vertical lines in Figure 2 indicate the period range 18 to 350 y corresponding to 
7 and 50 AU orbital separations from a 1 \msun star.  This model of the exoplanet
orbital distribution is very favorable for their detection  because the FOV of 
the GPI encompasses companions at the its peak.   

At the other extreme, we used power law relationship in period, $P^{-0.61}$, derived by
by Cumming et al. (2008) for planets detected by the radial velocity technique.   To bound
the power-law distribution we considered only planets at orbital periods of $log~P(d)= 2-6$, 
corresponding to orbital separations 0.42 to 194 AU.  We chose the inner bound because the
observed distribution of exoplanets discovered by the radial velocity technique departs
from a power-law at smaller separations (see for example Figs. 7 or 12 of Nielsen and Close
2009).  We chose the upper bound because Nielsen and Close found that hosts
with exoplanets distributed according to $P^{-0.61}$ cannot, with very high confidence,
harbor planets beyond 100 AU.   We binned C08's distribution at $\Delta log~P(d)=  1$.  

We evaluated the detectability of companions as follows.  For each BPMG member, we 
cycled through the possible companion masses listed in Tables 2 and 3. When using 
the DM91 period distribution, for each companion mass we generated 1000 orbits on 
which it might lie.  For the C08 period distribution, we generated 10000 orbits, 
all lying, in this case, between $log~P(d)= 2-6$.  The eccentricity distribution 
(Figure 3) is adapted from that observed for exoplanets as given on the web site 
of the Berkeley group\footnote{http://exoplanets.org/planets.shtml}.  
We binned the distribution by $\Delta e = 0.1$ and selected values of $e$ according to the
frequencies indicated in the figure.   

We assumed that all the other orbital parameters, most importantly the orbital inclination,
are distributed randomly.  With these values of the orbital parameters we calculated
the apparent positions of the 1000 (or 10000) model companions with respect to the host.  An example
for the BPMG member HIP 10680 and DM91 semi-major axis distribution is shown in Figure 4.
At its 39.4 pc distance, 378 of the model companions are hidden behind the occulting disc, 
403 fall outside the outer boundary of the FOV, leaving 219 within the FOV for possible 
detection.  Thus, at best, an imaging observation of HIP 10680 would start with only a 
$\sim 22\%$ chance of detecting a companion with the orbital parameters satisfying our assumed 
distributions.  For each companion mass a fresh set of randomly selected orbits and 
apparent positions are computed. The number of companions within the FOV therefore 
varies slightly according to the statistics of sampling.

To evaluate the detectability of the companions we consider the sensitivity of the GPI.
For each companion mass we obtain its apparent H magnitude at the distance of the
host, and evaluate the difference between it and the host.  For each of the hypothetical
companions that fall within the FOV at a particular angular separation from the host, 
we compare the 5-$\sigma$ sensitivity interpolated for the host star V magnitude
with the apparent magnitude difference.  If the difference is smaller than $\Delta$H
at 5-$\sigma$ we count the companion as a possible detection.  After evaluating all the
companions of a given mass within the FOV, we proceed to the next mass in Table 2 or 3.

\section {Results and Discussion}

\subsection {Detection of Companions in the BPMG}

The results of the modeling procedure for the hot and cold start brighness values and
DM91 and C08 orbital distributions are
displayed in Figures 5 to 8.  The figures show the fraction of 10 Myr-old possible companions 
detected at the 5-$\sigma$ level, at least, in 1 hr of GPI observation of the 15 
members of the BPMG (out of the 24 in Table 1), with V-band magnitudes within the senstivity 
limits of the GPI.  With declinations below $+15^\circ$ all 15 are observable by Gemini South.

For each BPMG member in Run 1, which use the hot start model exoplanets distributed according to the DM91
distribution,  the fraction of companions that could be detected  is approximately constant at 
$\sim$0.22 for companion masses $> 1~M_{J}$.  The small differences of the average detected 
fraction for a given host reflect the varying number of detectable model companions within the FOV at 
the host's distance and its apparent magnitude.  Fluctuations around the means are purely statistical
and are attributable to the different parameters selected for each trial companion mass (\S 2.2).  
For these companion masses, the star-companion contrast is within  GPI's 
sensitivity limit, and any companion in the FOV is detected.  Thus, for hosts
bright enough at V for good AO corection, the limiting factor in their detection is not the 
sensitivity of GPI but rather its FOV.  The small variations in the detection probability 
below $1~M_{J}$ arise in differences in host star brightness and distance.

At 9.9 pc distance, HIP 102409 is the closest of the of BPMG 15 members but the fraction
of its possible companions detected is comparable to that of the others.  At its distance,
the FOV samples the range 2-14 AU corresponding to a range of periods log P$\sim$3 to 4,
slightly below the peak of the assumed period distribution.   Evidently the increased
brightness of the host owing to its proximity is offset by the slightly smaller numbers of possible
companions lying within the FOV and illustrates the tradeoffs of distance, spectral type, and
apparent magnitude that affect the exoplanet detectability. 

The fraction of companions detected in Run 2 for cold start models with the DM91
distribution (Figure 6)  is much less than those in Run 1 because core-accretion
produces fainter companions.  For example, Tables 2 and 3 indicate that 10 $M_{J}$ 
exoplanets that formed from a hot start scenario have H-magnitudes of $\sim$11 to 12, 
while in  cold start models their H-band magnitude is $\sim$18.  
The fluctuations are statistical.  The detected fraction is not equal for all the hosts.
Again, HIP 102409 benefits from its close proximity 
to Earth.  Other stars such as HIP 215477, HIP 25486, and HIP 10679 also have
high detectability fractions primarily because they are relatively near the Earth.  While 
HIP 560, HIP 29964 and HIP 76629 have similar distances and H magnitudes, HIP 560 is much 
brighter in the V band which produces better AO correction and thus improved sensitivity
to exoplanets.

Results for the C08 power-law period orbital distribution are shown in Figures 7 and 8 (Runs
3 and 4).  Tthe fraction of detectable exoplanets drops dramatically to $\leq$0.05 
for companion masses  $> 1~M_{J}$.  Since we apply this distribution only to periods 
between $log~P(d)= 2$ to 6,  the fraction detected applies only to exoplanets in this range.  
The power-law distribution favors planets at short orbital periods and these tend to
fall behind GPI's occulting disk.  In these calculations, HIP 102409 has the highest 
fraction of exoplanet detectability; its proximity and the spectral type produce
a much more favorable star-planet contrast ($\Delta$H$\sim$13) than the other 
members ($\Delta$H$\sim$15-17).

\subsection{Discussion}

The essential result of our modeling is that the GPI has the potential
to image exoplanets with $M>1 M_{Jup}$ around host stars in the BPMG provided
(1) that the stars are sufficiently bright for the required AO correction, (2)
that the orbital distribution of exoplanets extends beyond $\sim 5$ AU, 
and (3) that the exoplanets are not much fainter than predicted by
the cold start formation model.  These qualifications are very important. 
The V-band brightness required for high quality AO correction provides the
severest cut of the BPMG sample, from the 29 members in Table 1 to the 15
observable stars, identified with an asterisk in Table 1 and appearing
in, for example, Fig. 5.

This exclusion also becomes
evident in a comparison of our analysis with imaging searches for exoplanets 
already  carried out (Masciadri et al. 2005; Biller et al. 2007).  
Nielsen and Close (2009) analyzed both surveys on a uniform  basis and 
provide a convenient list of the stars observed.   Biller et al. (2007) 
and Masciadri et al. (2005) observed 13 of the stars in Table 1 and did
not any detect exoplanets.  Only 3 of the 13 are sufficiently bright at V 
for the GPI (HIP 29964 SpTy K6/7, HIP 76629 K0, and V824 Ara G2).
Ten of the 13 stars observed by Masciadri et al and Biller et al are not accessible to GPI, 
one because it is too bright,  the A7 star HD 172555.  The remaining 9 stars, one
K7 and 8 early M's, are too faint for the GPI.  On the other hand, the 15 stars 
that are observable by GPI  include only one M star\footnote{It is unecessary 
therefore to include  a correction for the paucity of Jupiter-mass exoplanets 
associated with M stars (Johnson et al. 2007).}, the M1 HIP 102499  (and 
that only by  extrapolating the sensitivity curves to its V=9.9 mag).  
The others include 3 SpTy K stars, all  9 F and G stars in Table 1, and 
2 of the 4 A stars (the other 2 are too bright). Clearly, exoplanet 
searches in the BPMG using the GPI will have to favor the stars bright at V.

A comparison of Figs. 5 and 7, or Figs 6 and 8, illustrates the interplay of
GPI's FOV and the exoplanet orbital distribution.  Fortuitously, for targets in
the BPMG, the FOV includes the peak of the DM91 distribution.  The C08 distribution
decreases across the FOV as $a^{-0.92}$, a decrease of about a factor of 6 from
7 AU to 50 AU.  This accounts for the detectivity ratio $\sim 0.05/0.22$ for the
C08 and DM91 orbital distributions.  The ratio
is about the same for the hot start and cold start models but with greater
 scatter as a function of exoplanet mass and greater variablity among the 
stars in the cold start models.  Our use of the C08 distribution applies 
only to exoplanets with periods between $10^2$ and $10^6$ days; the ratio 
will be smaller if the C08 distribution applies at shorter periods.   
If the GPI is used to observe 15 stars in Figs. 5-8 and detects a few 
exoplanets, nominally 3, astronomers would learn that the DM91 distribution 
approximates the actual distribution at $7<a<50$ AU and that the planets 
formed according to the hot start scenario.  If on the other hand the search 
yields no exoplanet detections at all,  astronomers would be able to rule 
out the applicability of the DM91 at about the 95\% confidence
level, but would not be able to distinguish between the formation models.

Nielsen and Close's (2007) study of all the stars observed by Biller et al. (2007)
and Masciadri et al. (2005) shows that at the 68\% confidence level no more
than 20\% of the stars can have a $M>4 M_{Jup}$ companion in the host-exoplanet
separation range 22 to 807 AU for Baraffe et al's (2003) hot start models
and 25 to 557 AU for Fortney et al's (2008) cold start models. 
Nielsen and Close's result therefore suggests that for exoplanets 
with $M>4 M_{Jup}$ in this separation range the DM91 distribution may 
not apply.  Nielsen and Close also find that the observed
stars, with 95\% confidence limit, cannot harbor $M> 4M_{Jup}$ exoplanets
according to the C08 distribution beyond 65 AU (hot start models)
and 182 AU (cold start models).   These bounds are outside the GPI's
FOV.   If GPI observation of the 15 stars accessible in the BPMG were
to yield no detections at all, this would lower the upper bound on the
applicability of the C08 distribution.  It would also pose an interesting
puzzle given that exoplanets are associated with the A stars Fomalhaut
and HR 8799.  The non-detection would suggest that the exoplanet
orbital distribution depends on the mass of the host star and favors 
the presence of distant exoplanets if the star is sufficiently massive.

Figure 9 illustrates the dependence of exoplanet detectability on system age.
We considered a star like HIP 560, with its distance and spectral type, but of different
ages, 10 MY, 50 MY, 120 MY and 1 BY.   We used the hot start models and the DM91 distribution.
The result for 10 MY is the same in Fig. 5 and is representative of the age of the $\beta$
Pic and TW Hya moving groups.  The age of the AB Dor moving group lies between 50 and
120 MY.  The asymptotic detectability for cold start models and the C08 distribution
scales as in Figs. 5 to 8.  Figure 10 shows that the minimum detectable mass increases 
significantly with age and that the GPI will detect essentially  all stellar companions 
of nearby main sequence stars. At 1 BY, only companion masses greater than $70 M_{J}$  
are detectable by GPI.

\section{Summary}

We have found that for stars in the BPMG with $5<V<10$ mag,

1) GPI will be able to detect exoplanets with mass greater than $1 M_{J}$ for
both hot start or core-accretion scenarios. 

2) The V-band brightness for adequate AO correction requirement restricts the
target sample to stars earlier than $\sim$ K7 spectral type.

3) Given the GPI's high sensitivity, the two most important factors governing the 
exoplanet detection are the instrument's field of view and the period distribution 
of the exoplanets.  The number of exoplanets detected will distinguish between  
a distribution similar to that of the nearby F and G stars (i.e. DM91) and
the power-law distribution derived by Cumming et al. (2008).

4) Stars $\sim 10$ to 100 MY old are prime targets for GPI observation because of the favorable
star-planet contrast.  For older stars such as a 1-Byr F star at the 35 pc  typical distance of 
the BPMG, the  GPI will be able to detect all companions with $M> \sim 80 M_{J}$. 
  
\acknowledgments

We are grateful to the referee for many helpful suggestions that improved our paper and
thank Daniel Fabrycky and Jonathan Fortney for comments on the revised version.  We also thank James Graham 
for providing the current GPI sensitivity curves, Anand Sivaramakrishnan for discussions 
about the GPI and for advice about data fitting routines, and Josh Schlieder for providing 
the input data for Table 1.  T.K. thanks Dan Davis for a careful 
reading of TK's Honors thesis and comments.  T.K.'s work fulfilled requirements 
of an B.S. in Astronomy with Honors at Stony Brook University.  Our work was supported 
in part by NSF Grants AST 06-07612 and AST 09-07745.

\clearpage

\begin{deluxetable}{lcccccccccc}
\tabletypesize{\scriptsize}
\rotate
\tablecaption{$\beta$ Pic Moving Group members as of 9/2009 (adapted from Fern\'andez et al. 2008)\label{tbl-1}}
\tablewidth{0pt}
\tablehead{
\colhead{Name} & \colhead{RA} & \colhead{DEC} & \colhead{Spectral Type\tablenotemark{a}} & 
\colhead{$T_{eff}$\tablenotemark{b}} &
\colhead{$\msun$\tablenotemark{c}} & \colhead{D(pc)\tablenotemark{a}} & \colhead{V} &
\colhead{J\tablenotemark{d}} & \colhead{H\tablenotemark{d}} & \colhead{K\tablenotemark{d}} 
}
\startdata
HIP560$^*$	   & 00 06 50.09 & -23 06 27.1	& F2IV	  & 6890 & 1.7	& 39.1	& 6.181	 & 5.45	 & 5.331  & 5.24   \\
HIP215477$^*$  & 04 37 36.13 & -02 28 24.8	& F0V	  & 7200 & 1.8	& 29.8	& 5.216	 & 4.743 & 4.769  & 4.536  \\
\phm{---} GJ3305  & 04 37 37.47 & -02 29 28.4	& M0.5	  & 3850 & 0.6	& 29.8	& 10.74	 & 7.299 & 6.639  & 6.413  \\
HIP23309   & 05 00 47.13 & -57 15 25.5	& M0/1	  & 3785 & 0.5	& 26.3	& 10.112 & 7.095 & 6.429  & 6.244  \\
HIP25486$^*$   & 05 27 04.76 & -11 54 03.5	& F7	  & 6280 & 1.6	& 26.8	& 6.305	 & 5.267 & 5.086  & 4.926  \\
HIP27321   & 05 47 17.09 & -51 03 59.5	& A3V	  & 8720 & 2.0	& 19.3	& 3.853	 & 3.669 & 3.544  & 3.525  \\
HIP29964$^*$   & 06 18 28.21 & -72 02 41.5	& K6/7	  & 4133 & 0.9	& 38.5	& 9.995	 & 7.53	 & 6.984  & 6.814  \\
HIP76629$^*$   & 15 38 57.54 & -57 42 27.3	& K0V	  & 5250 & 1.5	& 39.8	& 8.16	 & 6.382 & 5.994  & 5.852  \\
HIP79881   & 16 18 17.90 & -28 36 50.5	& A0	  & 9520 & 2.2	& 43	& 4.782	 & 4.855 & 4.939  & 4.738  \\
V824 Ara$^*$   & 17 17 25.50 & -66 57 03.7	& G2IV	  & 5860 & 1.6	& 31.4	& 6.878	 & 5.288 & 4.907  & 4.701  \\
\phm{---} HD155555C  & 17 17 31.29 & -66 57 05.6	& (M4.5)  & 3310 & 0.2	& 31.4	& 12.7	 & 8.541 & 7.916  & 7.628  \\
HIP88399$^*$   & 18 03 03.41 & -51 38 56.4	& F5V	  & 6440 & 1.7	& 46.9	& 7.016	 & 6.158 & 6.022  & 5.913  \\
HIP88726$^*$   & 18 06 49.90 & -43 25 30.0	& A5V	  & 8200 & 1.9	& 43.9	& 5.67	 & 4.68	 & 4.488  & 4.386  \\
HD 172555  & 18 45 26.90 & -64 52 16.5	& A7	  & 7850 & 1.9	& 29.2	& 4.773	 & 4.381 & 4.251  & 4.298  \\
\phm{---} CD641208   & 18 45 37.03 & -64 51 45.9	& K7	  & 4060 & 0.8	& 29.2	& 11.028 & 6.906 & 6.318  & 6.096  \\
HIP92680$^*$   & 18 53 05.87 & -50 10 49.9	& K0Vp	  & 5250 & 1.5	& 49.7	& 8.414	 & 6.856 & 6.486  & 6.366  \\
nTel$^*$	   & 19 22 51.21 & -54 25 26.2	& A0Vn	  & 9520 & 2.2	& 47.7	& 5.02	 & 5.096 & 5.147  & 5.008  \\
HIP95270$^*$   & 19 22 58.94 & -54 32 17.0	& (F5.5)  & 6515 & 1.7	& 50.6	& 7.046	 & 6.2	 & 5.98	  & 5.91   \\
HIP102141  & 20 41 51.12 & -32 26 07.3	& (M4.5e) & 3305 & 0.2	& 10.2	& 11.02	 & 5.807 & 5.201  & 4.944  \\
HIP102409$^*$  & 20 45 09.53 & -31 20 27.2	& M1e	  & 3720 & 0.5	& 9.9	& 8.77	 & 5.435 & 4.831  & 4.528  \\
HD199143$^*$   & 20 55 47.67 & -17 06 51.0	& F8V	  & 6200 & 1.6	& 47.7	& 7.328	 & 6.206 & 5.945  & 5.81   \\
\phm{---} AZCap	   & 20 56 02.74 & -17 10 53.8	& (K7/M0) & 3955 & 0.7	& 47.7	& 10.631 & 7.849 & 7.249  & 7.039  \\
HIP10679$^*$   & 02 17 24.73 & +28 44 30.5	& G2V	  & 5860 & 1.6	& 34	& 7.75	 & 6.57	 & 6.355  & 6.262  \\
HIP10680$^*$   & 02 17 25.27 & +28 44 42.3	& F5V	  & 6440 & 1.7	& 39.4	& 7	 & 6.05	 & 5.84	  & 5.787  \\
HIP11437A  & 02 27 29.25 & +30 58 24.6	& K8	  & 3850 & 0.6	& 42.3	& 10.083 & 7.87	 & 7.235  & 7.08   \\
\phm{---} HIP11437B  & 02 27 28.09 & +30 58 40.1	& (K3)	  & 4730 & 1.3	& 42.3	& 12.46	 & 8.817 & 8.141  & 7.921  \\
HIP12545   & 02 41 25.89 & +05 59 18.4	& (M..)	  & 3850 & 0.6	& 40.5	& 10.2	 & 7.904 & 7.234  & 7.069  \\
HIP24418AB & 05 01 58.80 & +09 59 00.0	& M3V:	  & 3470 & 0.3	& 32.1	& 11.5	 & 7.212 & 6.657  & 6.37   \\
HIP112312A & 22 44 57.97 & -33 15 01.7	& M4	  & 3370 & 0.2	& 23.6	& 13.131 & 7.785 & 7.153  & 6.932  \\
\enddata

\tablenotetext{a}{Spectral types and distances were used from Table 7 of Fern\'andez et al. (2008).}
\tablenotetext{b}{Effective temperatures estimated from Kenyon and Hartmann's (1995) spectral type to effective temperature scale (see text}
\tablenotetext{c}{Masses estimated from Siess' online isochrone tool (see text)}
\tablenotetext{d}{J, H, and K magnitude values were used from the 2MASS survey (J.Schlieder, private communication.)}
\tablenotetext{e}{HIP11437A was given the temperature of an M0 star.}
\tablenotetext{f}{HIP11437B was given an approximate spectral class, K3, based on its V-I.}
\tablenotetext{g}{HIP12545 was given the approximate spectral class M0.}
\tablenotetext{*}{An asterisk indicates that star's V magnitude is in the range for adequate adaptive optics correction;
see Fig. 1 and \S 2.2.}

\end{deluxetable}

\clearpage

\begin{deluxetable}{lcc}
\tabletypesize{\scriptsize}
\tablecaption{Companion H magnitude at 10 Myr for hot start and core-accretion models (Fortney et al. 2008) \label{tbl-2}}
\tablewidth{0pt}
\tablehead{
\colhead{Mass ($M_{J}$)} & \colhead{$M_H$ hot start}  & \colhead{$M_H$ core-accretion} 
}
\startdata
1.0  & 18.27 & 19.43\\
2.0  & 16.05 & 18.11\\ 
4.0  & 13.95 & 18.52\\
6.0  & 12.82 & 18.73\\ 
8.0  & 12.13 & 18.79\\
10.0   & 11.52 & 18.69\\
\enddata
\end{deluxetable}

\clearpage

\begin{deluxetable}{lc}
\tabletypesize{\scriptsize}
\tablecaption{Companion H magnitude at 10 Myr (Baraffe et al. 2003)\label{tbl-3}}
\tablewidth{0pt}
\tablehead{
\colhead{Mass ($M_{J}$)} & \colhead{$M_H$}
}
\startdata
0.5 & 20.16 \\
1.0  & 17.79 \\
2.0  & 15.79 \\ 
3.0  & 14.65 \\
4.0  & 13.85 \\
5.0  & 13.22 \\
6.0  & 12.68 \\ 
7.0  & 12.27 \\
8.0  & 11.87 \\
9.0  & 11.54 \\
10.0   & 11.26 \\
12.0  & 10.75 \\
15.0  & 9.87  \\
20.0   & 8.91  \\
30.0   & 8.28  \\
40.0   & 8.25  \\
50.0   & 7.84  \\
60.0   & 7.73  \\
70.0   & 7.52  \\
80.0   & 7.32  \\
90.0   & 7.14  \\
100.0    & 6.93  \\
\enddata
\end{deluxetable}

\clearpage




\begin{figure}
\epsscale{.80}
\plotone{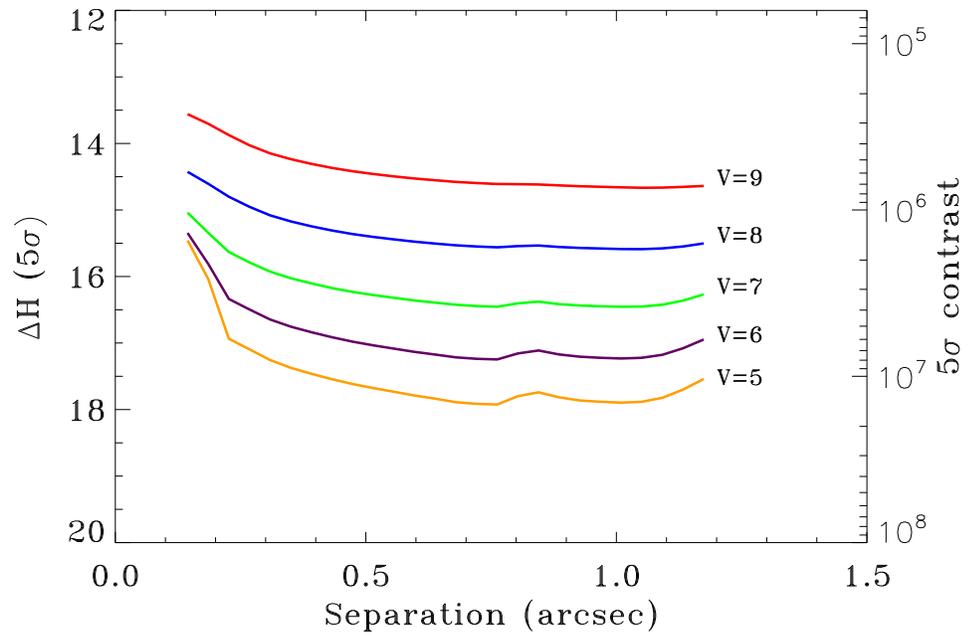}
\caption{Detectable H magnitude difference in 1 hour of GPI observation, plotted as a function of 
the angular separation between the companion and host star.  (Graham, private comm.).  
The curves are given at 5 values of the host star brightness, V= 5 to 9 mag.}
\label{sens_plot}
\end{figure}

\clearpage

\begin{figure}
\epsscale{.80}
\plotone{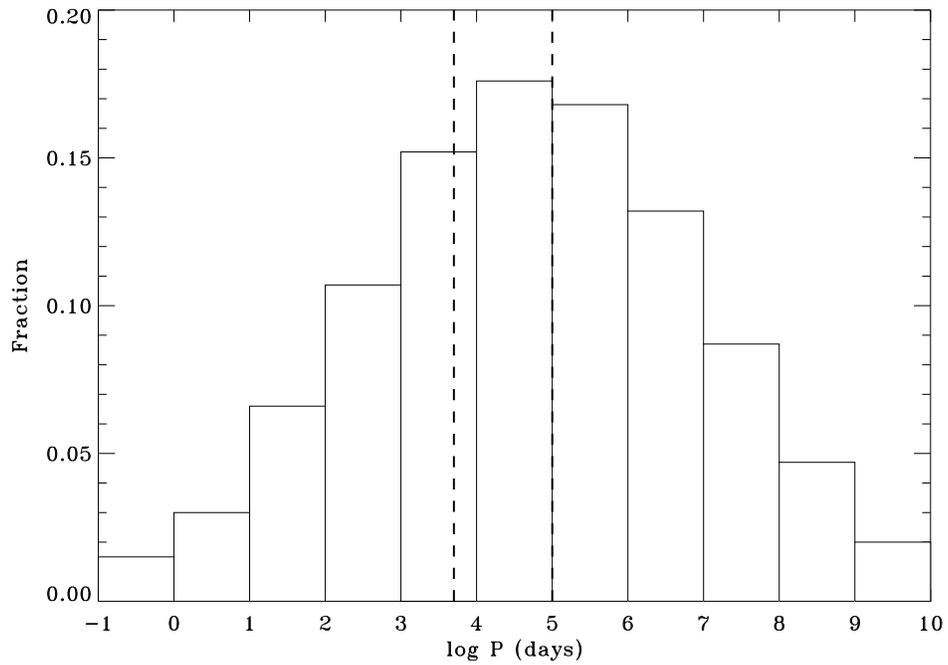}
\caption{Period distribution used in the modeling routine, adapted from Duquennoy and Mayor (1991). The dotted 
lines indicate the period range encompassed by the GPI's FOV for a 1 \msun ~ star at the median 
distance of the BPMG, 35 pc.}
\label{p_plot}
\end{figure}

\clearpage

\begin{figure}
\epsscale{.80}
\plotone{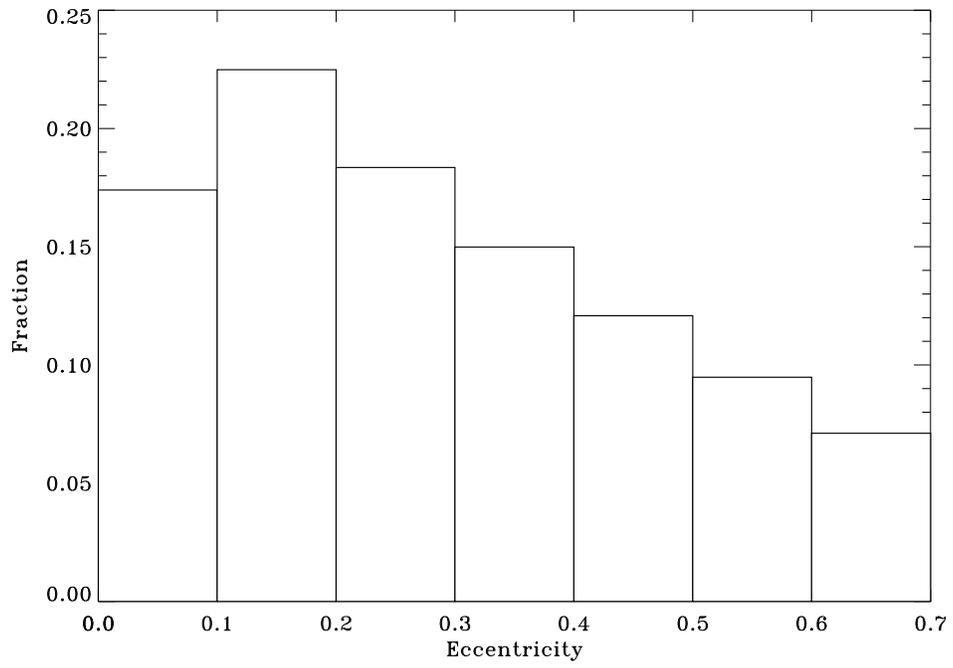}
\caption{Eccentricity distribution used in the modeling routine, adapted from exoplanets.org.}
\label{ecc_plot}
\end{figure}

\clearpage

\begin{figure}
\epsscale{.80}
\plotone{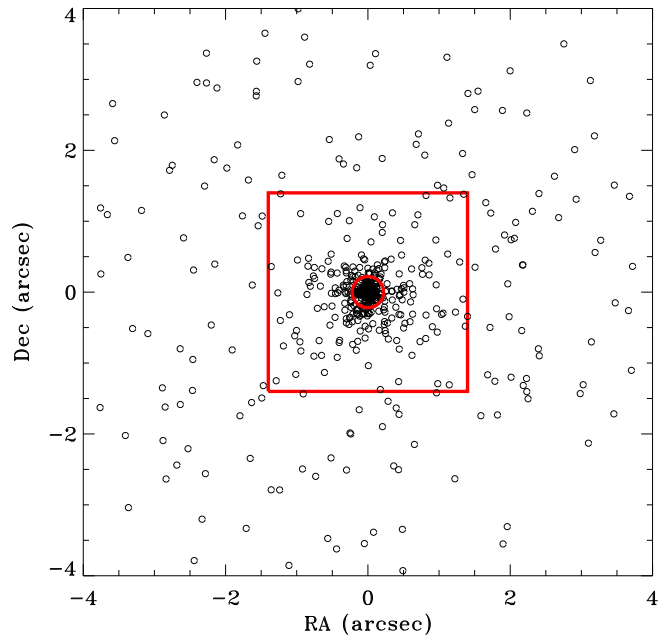}
\caption{Sample output of 1000 companions to HIP10680, plotted as RA and Dec separation 
with respect to the host. This calculation used the Duquennoy and Mayor (1991) orbital distribution as in Fig. 2.
 Of the 1000 companions,  378 fall behind
the occulter, 403 fall outside of the FOV, leaving 219 within the FOV of GPI (see text).} 
\label{fov_plot}
\end{figure}

\clearpage

\begin{figure}
\begin{center}
\includegraphics[scale=0.8]{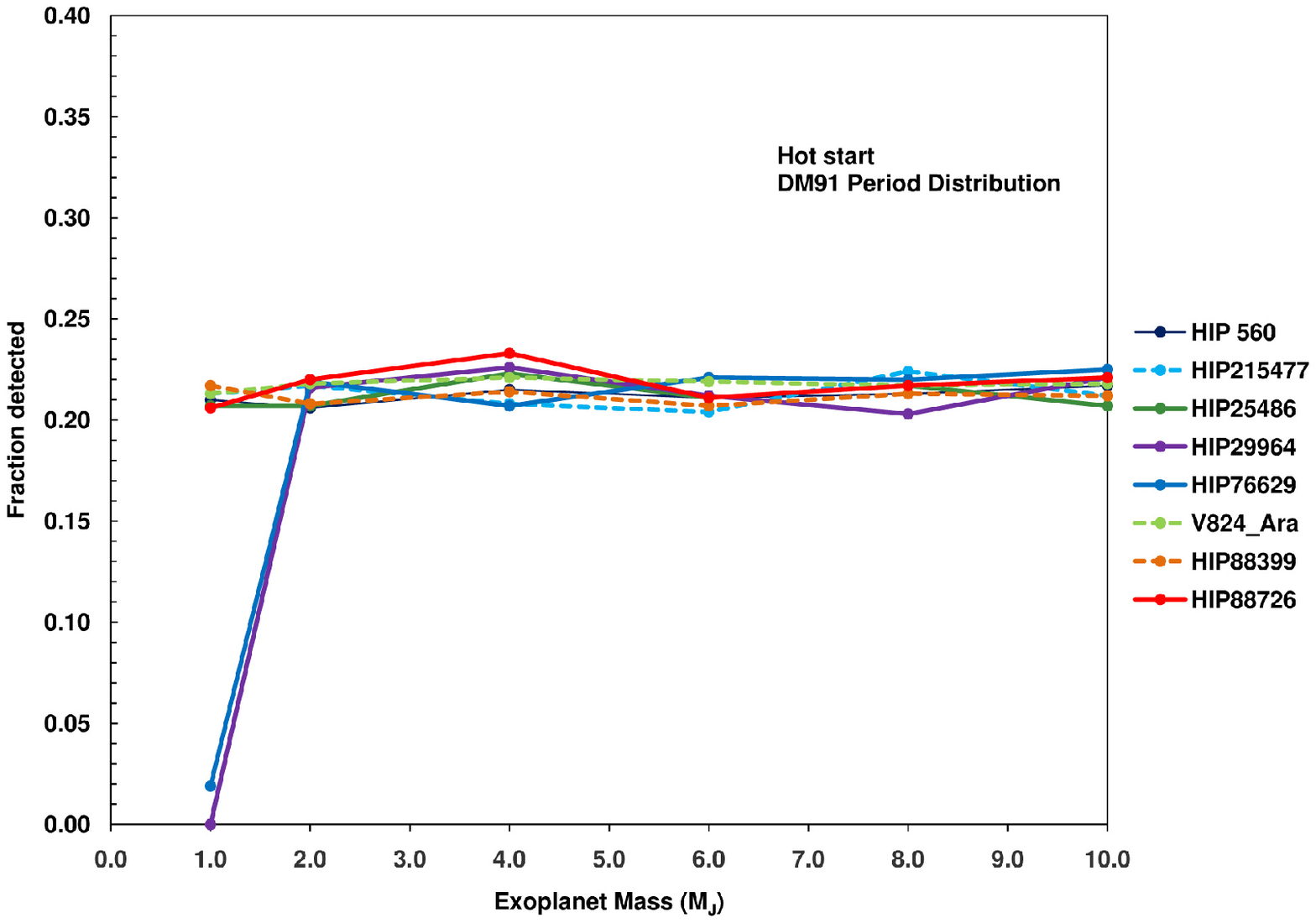}
\includegraphics[scale=0.8]{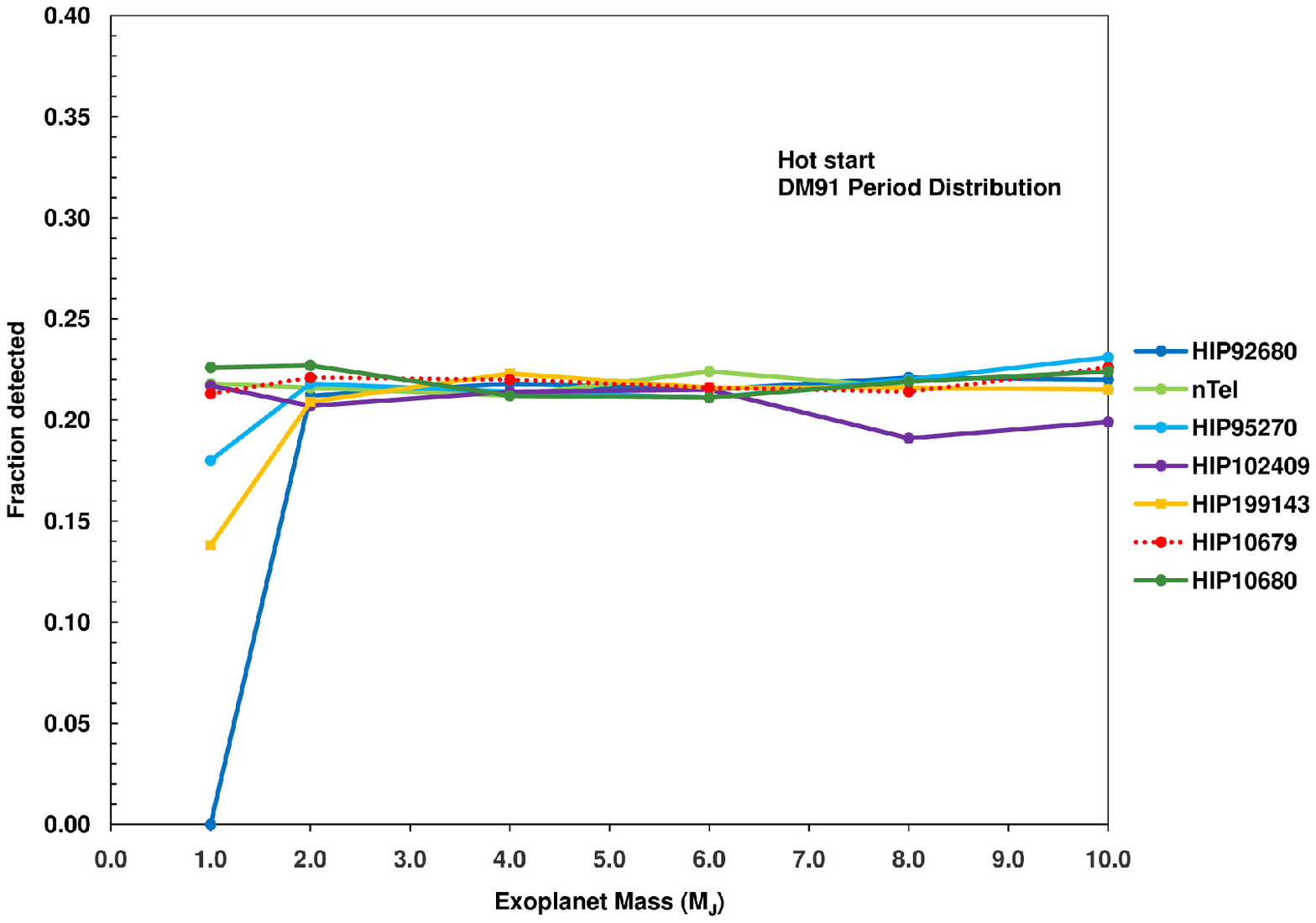}
\end{center}
\caption{Fraction of possible companions detected in 1 hr of GPI observation with at least 
$5-\sigma$ confidence as a function of companion mass, using the hot start model and DM91 period distribution (Run 1).}
\label{res1}
\end{figure}

\clearpage

\begin{figure}
\begin{center}
\includegraphics[scale=0.8]{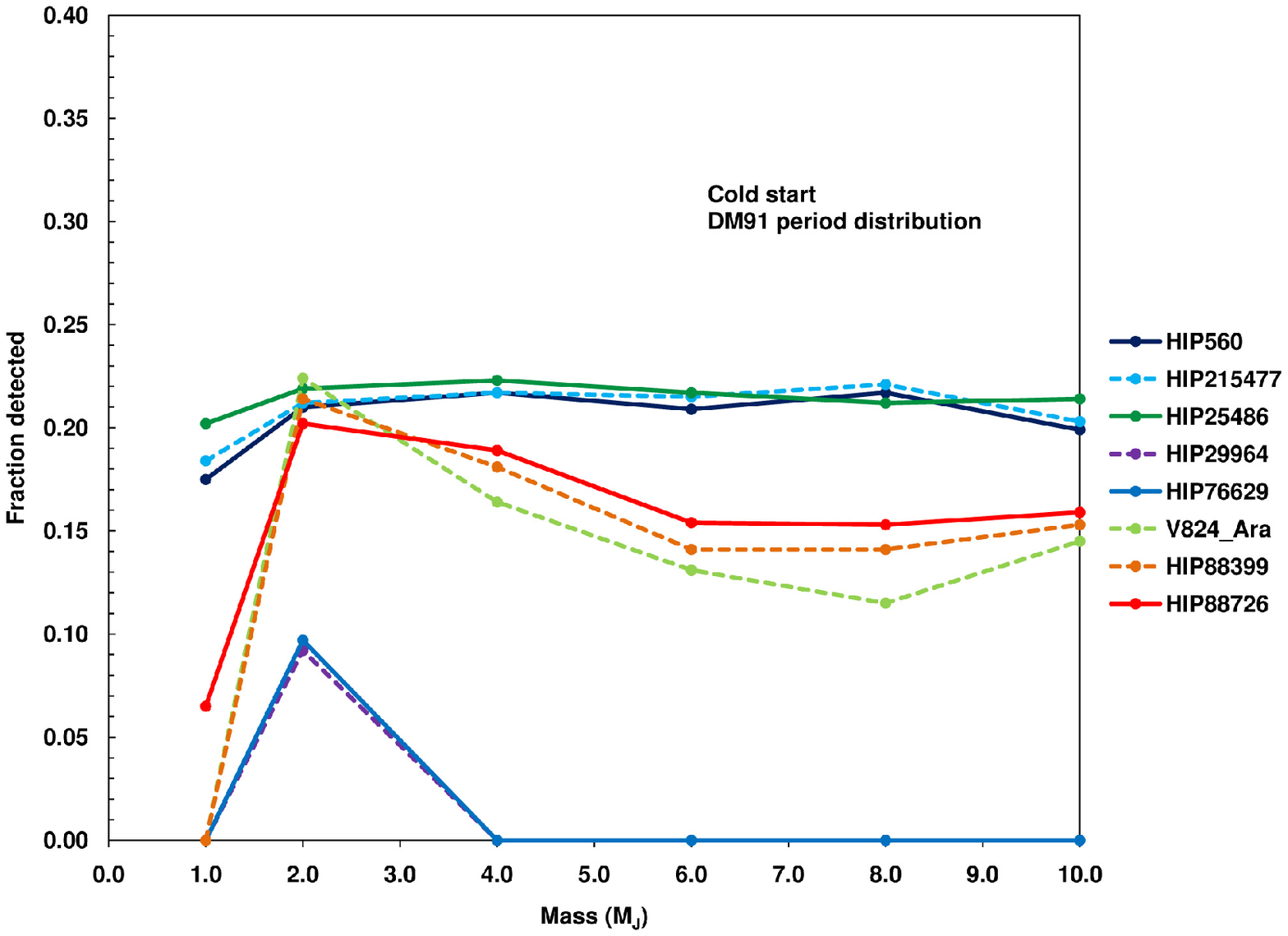}
\includegraphics[scale=0.8]{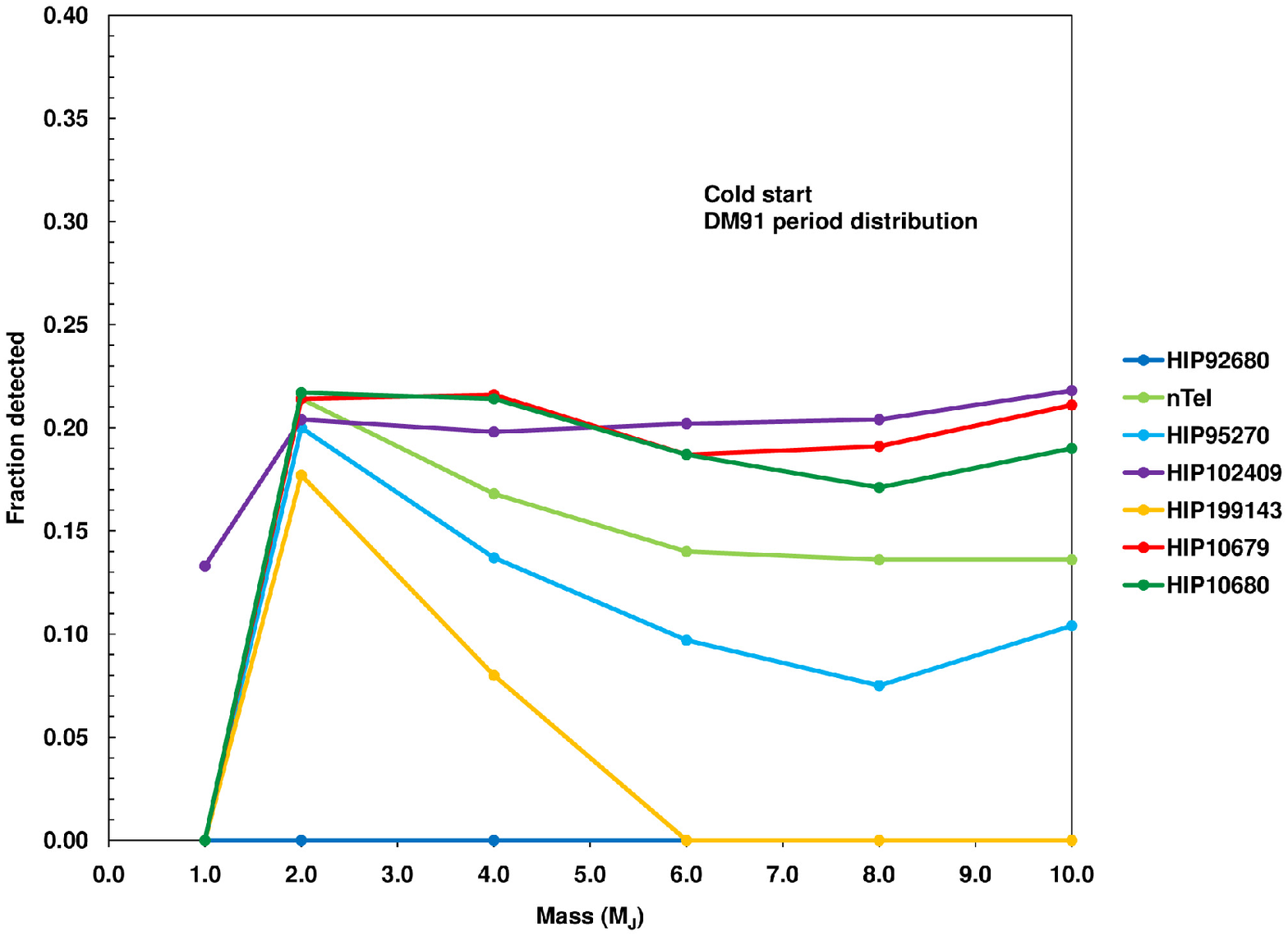}
\end{center}
\caption{Fraction of possible companions detected in 1 hr of GPI observation with at least 
$5-\sigma$ confidence as a function of companion mass, using the cold start model and DM91 period distribution (Run 2).}
\label{res2}
\end{figure}

\clearpage

\begin{figure}
\begin{center}
\includegraphics[scale=0.8]{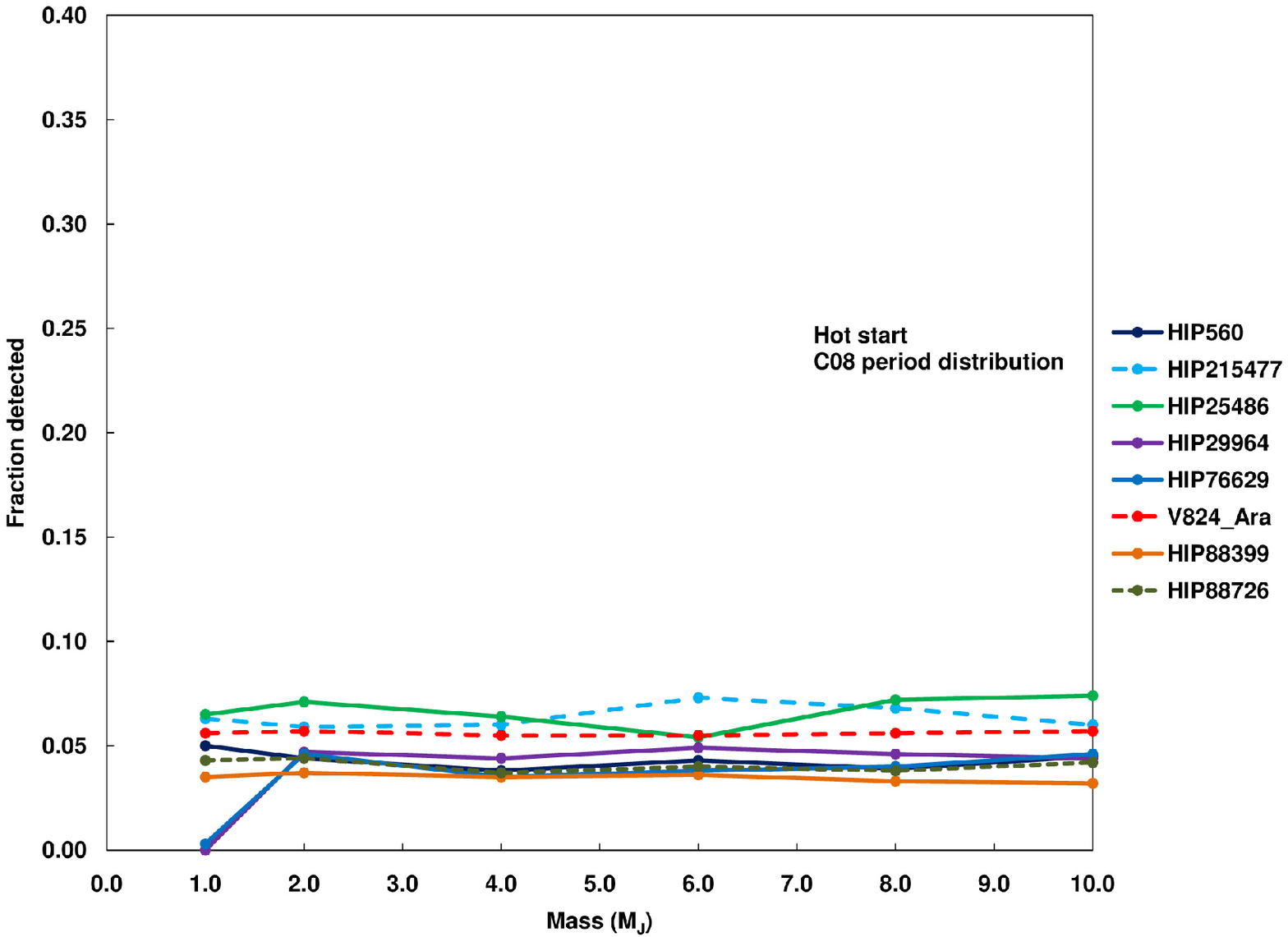}
\includegraphics[scale=0.8]{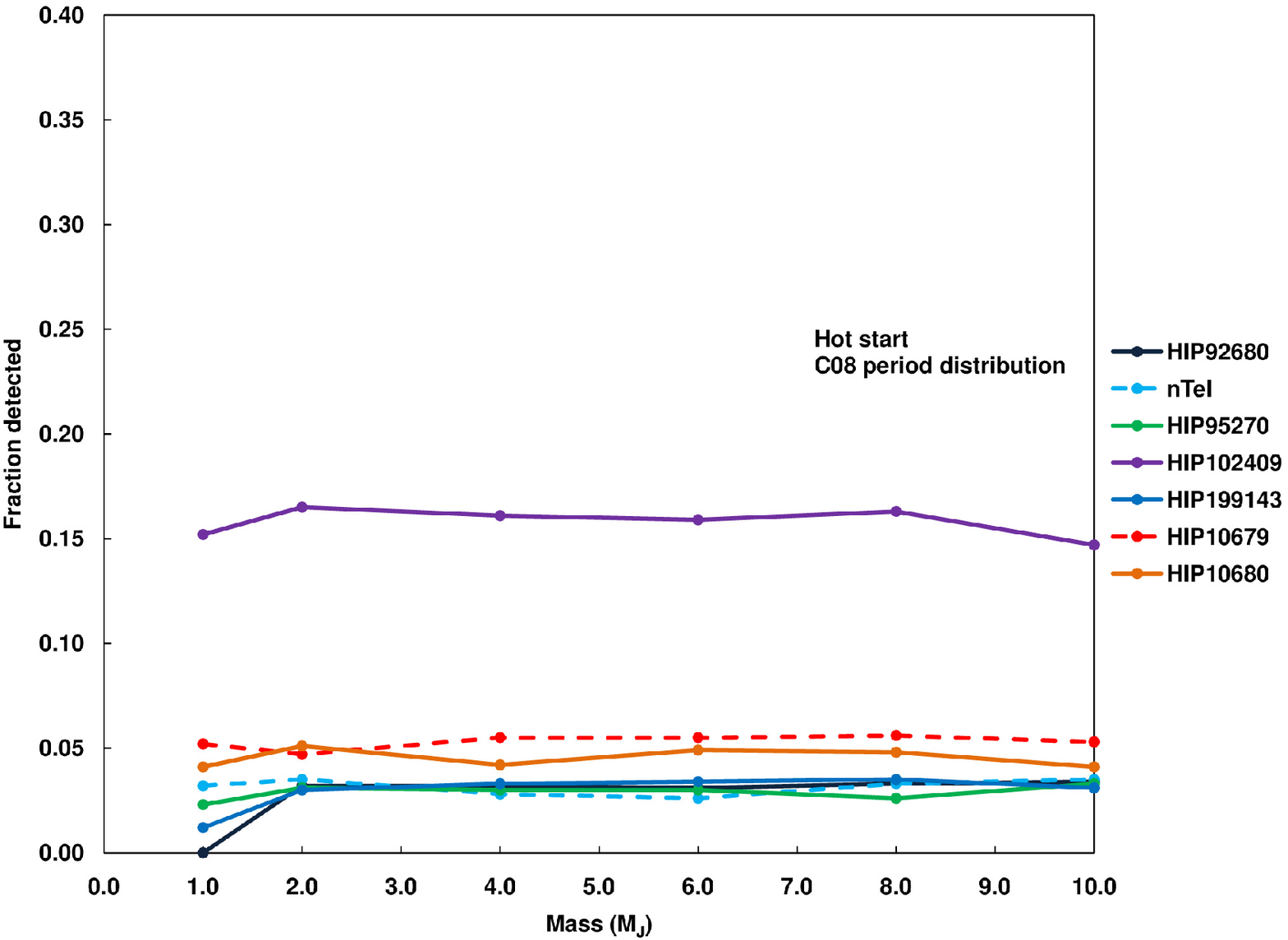}
\end{center}
\caption{Fraction of possible companions detected in 1 hr of GPI observation with at least 
$5-\sigma$ confidence as a function of companion mass, using the hot start model and C08 period distribution (Run 3).}
\label{res3}
\end{figure}

\clearpage

\begin{figure}
\begin{center}
\includegraphics[scale=0.8]{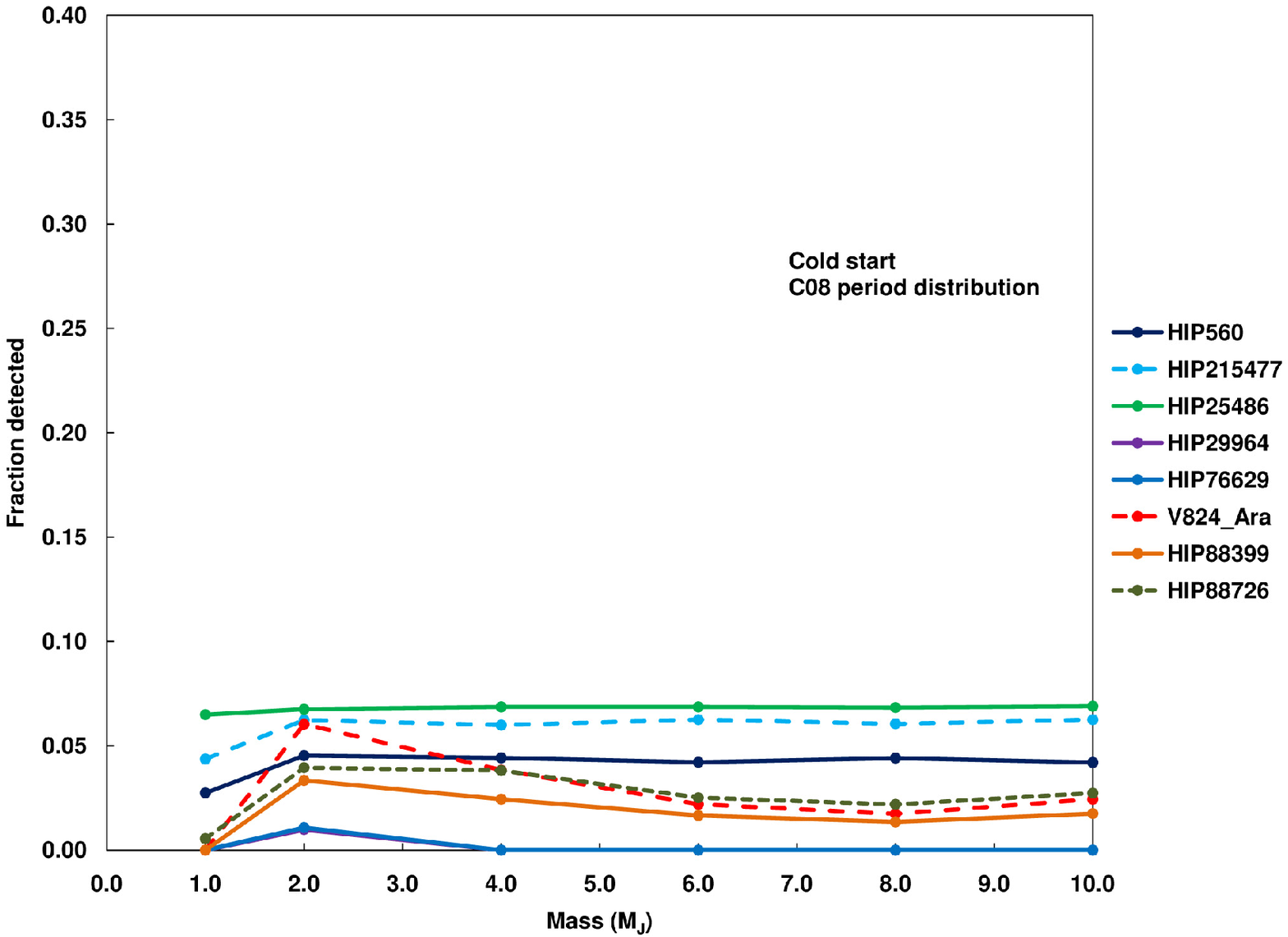}
\includegraphics[scale=0.8]{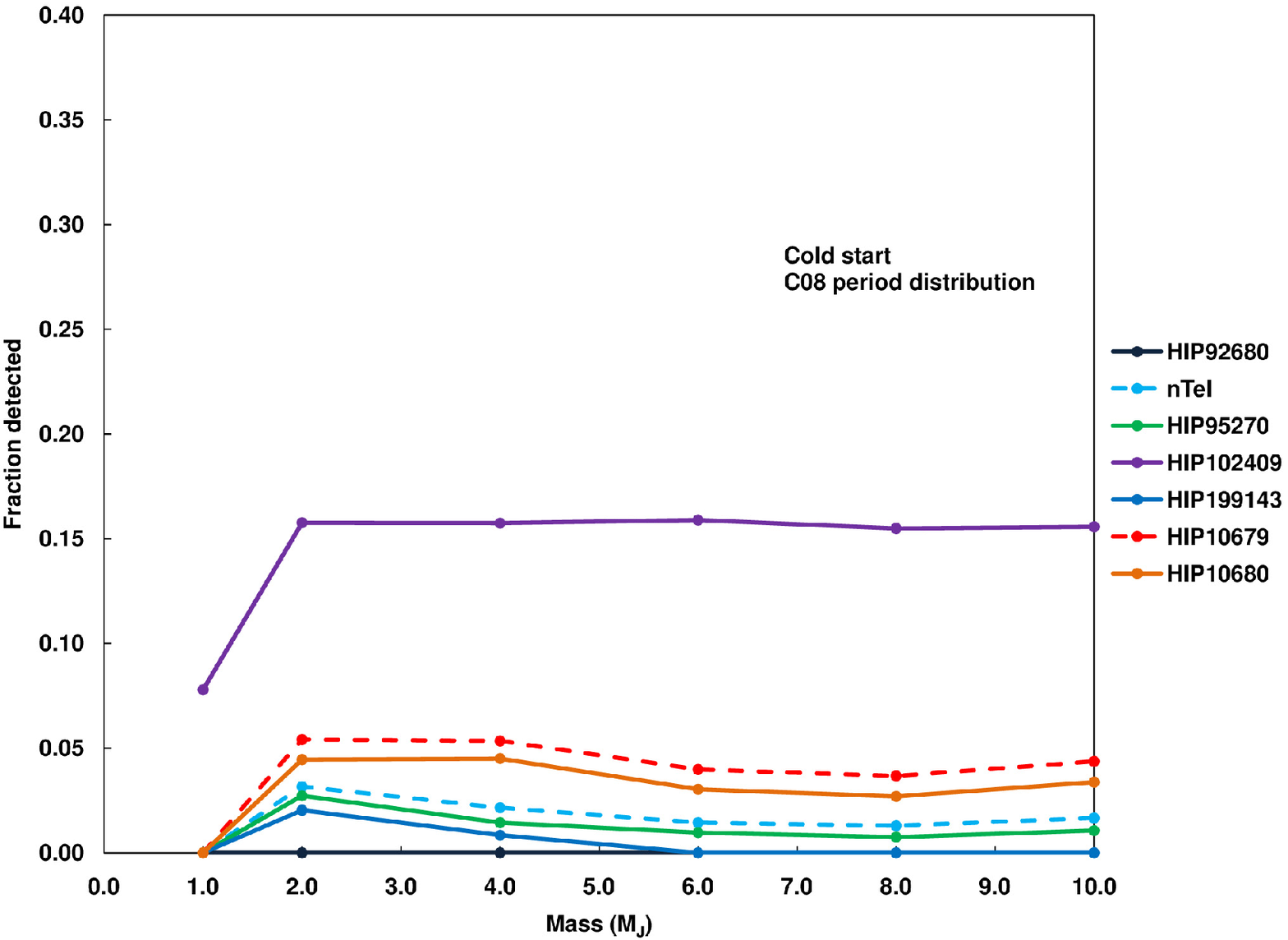}
\end{center}
\caption{Fraction of possible companions detected in 1 hr of GPI observation with at least 
$5-\sigma$ confidence as a function of companion mass, using the cold start model and C08 period distribution (Run 4).}
\label{res4}
\end{figure}

\clearpage

\begin{figure}
\includegraphics[scale=0.6]{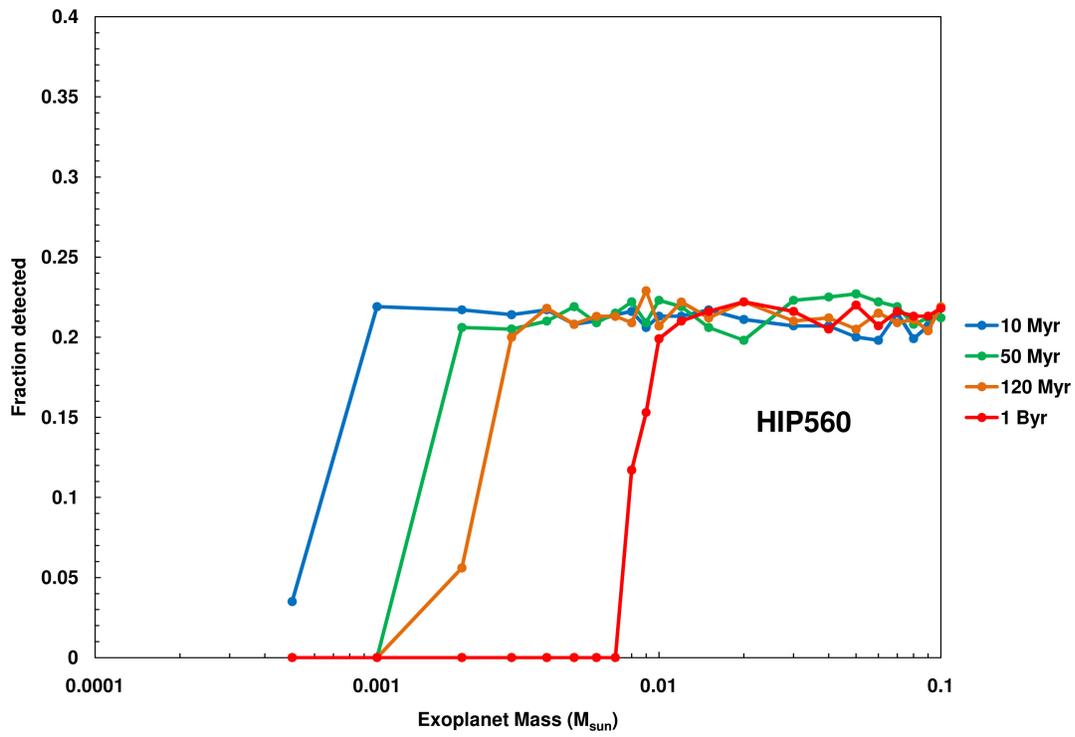}
\caption{Fraction of possible companions detected (as in Figure 5), but for a star like
HIP560 with varying ages: 10 Myr (close to the actual age), 50 Myr, 120 Myr, and 1 Byr.}
\label{res6}
\end{figure}

\clearpage

\end{document}